\documentclass[structabstract]{aa}  
\usepackage{graphicx}
\usepackage{natbib}
\usepackage{subfigure}
\usepackage{pdflscape}

\usepackage{caption} 
\usepackage{subfig}
\usepackage{verbatim}
\usepackage{pdfpages}
\usepackage[figuresright]{rotating}
\usepackage[justification=raggedright]{caption}
\bibpunct{(}{)}{;}{a}{}{,}

\usepackage{txfonts}
\begin{document}

   \title{X-ray deficiency on strongly accreting T Tauri stars.}
\subtitle{Comparing Orion with Taurus}
     
   \author{I. Bustamante \inst{1,2,3}
   \and B. Mer\'in \inst{1}
   \and H. Bouy \inst{2}
   \and C. Manara \inst{4}
   \and \'A. Ribas \inst{1,2,3}
   \and P. Riviere-Marichalar \inst{1}
               }

   \institute{European Space Astronomy Centre (ESA), P.O. Box, 78, 28691 Villanueva de la Ca\~{n}ada, Madrid, Spain
   \and Centro de Astrobiolog\'ia, INTA-CSIC, P.O. Box - Apdo. de correos 78, Villanueva de la Ca\~{n}ada Madrid 28691, Spain
   \and ISDEFE - ESAC, P.O. Box, 78, 28691 Villanueva de la Ca\~{n}ada, Madrid, Spain
  \and Scientific Support Office, Directorate of Science and Robotic Exploration, European Space Research and Technology Centre (ESA/ESTEC), Keplerlaan 1, 2201 AZ Noordwijk, The Netherlands 
   }
      \date{Received 21/10/2015; accepted 04/11/2015}

% \abstract{}{}{}{}{} 
% 5 {} token are mandatory
 
  \abstract
   {Depending on whether a T Tauri star accretes material from its circumstellar disk or not, different X-ray emission properties can be found. The accretion shocks produce cool heating of the plasma, contributing to the soft X-ray emission from the star.}
   {Using X-ray data from the \textit{Chandra Orion Ultra-deep Project} and accretion rates that were obtained with the \textit{Hubble Space Telescope}/WFPC2 photometric measurements in the Orion Nebula Cluster, we studied the relation between the accretion processes and the X-ray emissions of a coherent sample of T Tauri sources in the region.}
   {We performed regression and correlation analyses of our sample of T Tauri stars between the X-ray parameters, stellar properties, and the accretion measurements.}
  {We find that a clear anti-correlation is present between the residual X-ray luminosity and the accretion rates in our samples in Orion that is consistent with that found on the \textit{XMM-Newton Extended Survey of the Taurus molecular cloud} (XEST) study. A considerable number of classified non-accreting sources show accretion rates comparable to those of classical T Tauri Stars (CTTS). Our data do not allow us to confirm the classification between classical and weak-line T Tauri stars (WTTS), and the number of WTTS in this work is small compared to the complete samples. Thus, we have used the entire samples as accretors in our analysis. We provide a catalog with X-ray luminosities (corrected from distance) and accretion measurements of an Orion Nebula Cluster (ONC) T Tauri stars sample.}
{Although Orion and Taurus display strong differences in their properties (total gas and dust mass, star density, strong irradiation from massive stars), we find that a similar relation between the residual X-ray emission and accretion rate is present in the Taurus molecular cloud and in the 
accreting samples from the Orion Nebula Cluster. The spread in the data suggests dependencies of the accretion rates and the X-ray luminosities other than the stellar mass, but the similarity between Orion and Taurus hints at the environment  not being one of them. The anti-correlation between the residual X-ray luminosity and mass accretion rate is inherent to the T Tauri stars in general, independent of their birthplace and environment, and intrinsic to early stellar evolution.}


 \maketitle
%
%________________________________________________________________

\section{Introduction}
 The interaction between the central star and its surrounding disk in young stellar objects (YSOs) could be one the most important processes in the evolution of young stars, proto-planetary disks, and planetary system formation but, paradoxically, one of the least understood. Ultraviolet and X-ray irradiation from the star affects the dispersal and evolution of its surroundings \citep{Owen2010}, and the magnetic activity of the star has an important role in the mass-accretion rate from the disk onto the central object. The X-ray emission from the central star should photo-ionize the circumstellar material \citep{Gorti2009}, affecting its chemical composition, accreting processes \citep{Alexander2014}, outflows, and potential planetary atmospheres. 

T Tauri stars present X-ray luminosities ($L_X$) near the saturation limit (log($L_X$ /$L_{bol}$) = -􏰅3) found for main-sequence (MS) stars, $L_X/L_{bol}$ $\sim$ $10^{-3.5}$, i.e., sometimes orders of magnitude higher than more evolved stars. For MS stars the X-ray activity is highly related with the rotation period, a relation given by the power law $L_X$/$L_{bol}$ $\propto$ $P_{rot}^{-2.6}$ \citep{Pizzolato2003}. But this relation has not been found for T Tauri stars \citep{Preibisch2005origin}. Several reasons have been postulated: the totally convective nature of young stars could be generating a different magnetic activity than that of MS stars (solar-like dynamo); a possible star magnetic field that is coupling with its surrounding disk; or the X-ray emission from accretion shocks could be altering the total amount of X-ray luminosity \citep{Preibisch2005origin}. 

In the accretion process, material from the disk falls onto the stellar photosphere, generating a characteristic shock spectrum and an excess emission ($L_{acc}$), which can be measured with spectroscopic or photometric methods \citep{Calvet1998}. This accretion luminosity can be linked to the mass accretion rate, $\dot{M}_{acc}$, through the relation: $\dot{M}_{acc}$ = $L_{acc}R_{*}$/$0.8GM_{*}$, with $R_{*,}$ and $M_{*}$ being the radius and the mass of the star, respectively, and 0.8 being the factor that accounts for the assumption that the infall originates at a magnetospheric radius of $R_{m}$ = 5 \citep{Gullbring1998}. The excess emission is mostly seen in the blue part of the spectrum (\textit{U-band} excess) and in the emission lines ($H\alpha$ equivalent width) \citep{Hartmann1998}.

The accretion processes were thought to play an important role in the X-ray activity of these stars, although several studies have found that strong accretors present lower $L_X$ \citep{Preibisch2005origin}. Different explanations have been elaborated, such as disk opaqueness to certain $L_X$ energies \citep{Gudel2007XESTb}, thermalization and absorption of the soft component of the X-ray luminosity ($\sim$5 keV) \citep{Calvet1998} or changes in the coronal magnetic field activity provoked by the accretion processes \citep{Telleschi2007}. Also, a higher magnetic activity of the young star would result in highly energetic flares, showing large enough peaks in the X-ray measurements  to even connect the inner disk to the star's photosphere \citep{Aarnio2010}.

Several studies have been carried out regarding the relation between mass accretion rates and X-ray emission. One of these is the anti-correlation found between the residual $L_X$ and the $\dot{M}_{acc}$ by \cite{Telleschi2007}, using data part of the \textit{XMM-Newton Extended Survey of the Taurus molecular cloud} (XEST). Our main objective is to compare the mass accretion rates -- X-ray emission relation found in that article \citep{Telleschi2007} for the low mass T-association of the Taurus Molecular Cloud (TMC) with data on the Orion Nebula Cluster (ONC), and to look for any possible environmental effect. We used the X-ray emission data from the \textit{Chandra Orion Ultra-deep Project} (COUP) \citep{Getman2005observations, Preibisch2005origin} and the accretion measurements from \cite{Manara2012}, which use data from the \textit{Hubble Space Telescope} (\textit{HST}) Treasury Program on the Orion Nebula Cluster \citep{Robberto2013}. 

This article is organized as follows: Sect. ~\ref{sec:observations} describes the different datasets we use in our study and the cross-matching procedure. Sect. ~\ref{sec:analysis} illustrates the selection of the sub-samples we use. In Sect. ~\ref{sec:correlations} we explain the relation between X-ray emission and accretion processes we want to analyze and the correlation results we obtained, which are fully discussed in Sect. ~\ref{sec:discussion}. In Section ~\ref{sec:conclussions}, we summarize the conclusions of this work.

\section{X-rays and accretion in the ONC}
\label{sec:observations}

\subsection{Taurus and Orion comparison}
The TMC is the nearest largest \citep[distance $\sim$140 pc, ][]{Loinard2005} star-forming region known. It is composed of several loosely associated molecular clouds, each of them producing small numbers of YSOs. Its small star density (1-10 stars $pc^{-2}$, \citep{Gomez1993}) prevents a notable influence from gravitational effects, outflows, or strong UV radiation fields, which creates a calm evolutionary environment, compared to other star-forming regions. 

Although several other X-ray studies have been carried out in the region, the XEST project provided the most sensitive and complete sample of X-ray detections \citep{Telleschi2007, Gudel2007XESTa}. The X-ray emission of TTS is usually between $L_X$ $\sim$ $10^{29}$ - $10^{31}$ erg $s^{-1}$, clearly higher than the instrument-detection limit of $L_X$ $\sim$ $10^{28}$ erg $s^{-1}$, for typical integration times. In fact, the XEST presents X-ray detection statistics consistent with near-completeness: 126 out of 152 TMC members are detected in X-rays \citep{Gudel2007XESTa}. Out of a this sample, 55 are classified as strong accretors (or classical T Tauri stars) and 45 as weak accretors (or weak-line T Tauri stars), a classification that is based on the equivalent width of the H$\alpha$ line. We refer to \cite{Telleschi2007} for a complete description of their sample selection and analyses.

In contrast, the ONC presents a much denser picture, with a distance of $\sim$414 pc \citep{Menten2007}. This massive cluster of young stars ($\sim$ $10^6$ yr) located in the Orion Molecular Cloud is illuminated by two O-type stars ($\Theta^1$ Ori C and $\Theta^2$ Ori A), has a star population of $\sim$1600 optically visible sources, and is recognized as a benchmark laboratory for star and planet formation, and the closest massive cluster to the Sun. It represents a coherent and homogeneous group of pre-main sequence (PMS) stars \citep{Hillenbrand1997, DaRio2010, DaRio2012}, and a perfect  bench test for a comparison with the TMC.
Table \ref{tab:tmc_onc_comparison} summarizes some key characteristics of both regions.

\subsection{X-ray emission in the ONC - COUP}
We use the data available from the COUP \citep{Preibisch2005origin, Feigelson2005, Getman2005membership}, or more precisely, from the \textit{Chandra} Orion Ultra-deep Point Source Catalog (http://heasarc.gsfc.nasa.gov/W3Browse/chandra/coup.html). 

The COUP was a deep observation of the ONC using the Advanced CCD Imaging Spectrometer (ACIS) mounted on the \textit{Chandra} X-ray telescope. This was a continuous 838 ks exposure program over a period of 13.2 days in January 2003. It detected 1 616 individual X-ray sources, with typical positional uncertainties of $<$0."3. This observation of the ONC \citep{Getman2005observations}  resulted in the detection of $\sim$1400 X-ray-emitting PMS and is the richest source of X-ray data ever obtained for the ONC. We refer to \cite{Getman2005membership, Getman2005observations} for a full review of the program. 

\subsection{Accretion data}
For the accretion measurements, we used the data of \cite{Manara2012} on the ONC, in which they use the photometric catalog from the \textit{HST} Treasury Program of \cite{Robberto2013}. This program observed the ONC for 104 HST orbits of \textit{HST} time with the Advanced Camera for Surveys (ACS), the Wide-Field/Planetary Camera 2 (WFCP2), and the Near-Infrared Camera and Multi-Object Spectrograph (NICMOS), using 11 filters in total that range from the \textit{U} to the \textit{H} band. With the data obtained from WFPC2, \cite{Manara2012} calculate the mass accretion rates of almost $\sim$600 PMS stars. Here we summarize the key aspects of their work and briefly discuss  the two methodologies they use to determine the accretion rates. In the following, we divide the discussion on this dataset using two subsamples that were created by the method  used to determine the accretion rates: the \textit{U-band} excess and the $H{\alpha}$ line flux. For a complete analysis of the data selection and procedure, see their article. 

\cite{Manara2012} construct a two-color diagram (2CD) with the \textit{UBI} photometric measurements, and simultaneously obtain an estimate of the extinction, \textit{A$_{V}$}, and accretion luminosity, $L_{acc}$/L$_{tot}$. They assume that the displacement of the observed sources from the theoretical isochrone on the 2CD is caused by a combination of these two processes. They assume spectral types from the literature for all their targets. 

\textit{U-band excess} - From the 2CD that they converted they derive $L_{acc}$/$L_{tot}$ into $L_{acc}$/$L_\odot, $ taking into consideration the stellar luminosities with the relation $L_{acc}$/$L_\odot$ = ($L_{acc}$/$L_{tot}$)/(1-$L_{acc}$/$L_{tot}$) * $L_*$/$L_\odot$ ($L_*$ being the stellar luminosity).

\textit{$H\alpha$ line} -  \cite{Manara2012} also use the $H\alpha$ luminosity as an accretion tracer. With the $H\alpha$ photometry available from the WFPC2 catalog (more precisely, the photometry from the \textit{F656N} filter), they measure the excess with respect to the photospheric ($H\alpha$ - I) color for each star, which is dependent on $T_{eff}$. They then convert this $H\alpha$ excess into equivalent width. They consider the sources that fulfill the condition 3 \AA\ $<$ EW$_{H\alpha}$ $<$ 1000 \AA\ as accretors. Finally, they derive the accretion luminosity $L_{acc}$ from the $H{\alpha}$ luminosity $L_{H{\alpha}}$ with the relation log($L_{acc}$/$L_\odot$) = (1.31 $\pm$ 0.03)log($L_{H{\alpha}}$/$L_\odot$)+(2.63 $\pm$ 0.13). The $L_{acc} - L_{H\alpha}$ relation was calibrated using the \textit{U-excess} data. The masses of the stars were computed with an evolutionary model interpolation of the position of them on the Hertzprung-Russel diagram (HRD).

\cite{Manara2012} obtain accretion rates for 244 objects using the \textit{U-excess} method, and for 486 using the $H\alpha$ one. These subsamples are independent of one another. The excess in the U-band is a direct proxy of accretion, while the $L_{H\alpha}$ is an indirect one. Thus, the accretion rate measurements of the  \textit{U-excess} subsample are more reliable than those of the $H\alpha$ one.

\cite{Manara2012} discard close binaries and visual proplyds \citep{Ricci2008} from their sample. Thus, our subsamples are cleansed of these objects too. Consequently, our catalogs are  only composed of young accreting stars. We will refer to these catalogs as the \textit{U-excess} and the $H\alpha$ subsamples, respectively.

\subsection{Cross-matching}
The first step in our analysis was to find the COUP X-ray counterparts for the sources in the \textit{HST} sub-samples. Using the nearest neighbors method, we made a first cross-match with the sky positions of the sources of the COUP catalog and those of the \textit{Manara2012} subsamples, using a matching distance of 1" as a first approximation. We chose this distance taking into account the astrometric accuracy of \textit{Chandra}: the 99\% limit on positional accuracy is 0.8", in the worst case being an offset of 1.1" \footnote{http://cxc.harvard.edu/cal/ASPECT/celmon/}. With this matching distance, we obtained 170 pairings with the \textit{U-excess} group, and 322 pairings with the $H\alpha$ one. 

\begin{figure*}[htbp]
 \begin{center}
    \includegraphics[width=0.5\textwidth]{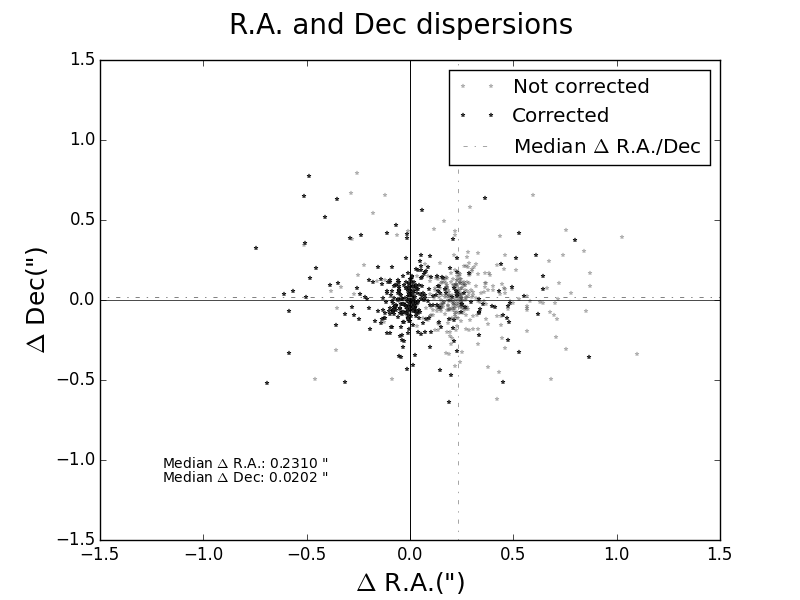}
 \caption{Right ascension and declination offsets between the \textit{HST} $H{\alpha}$ sources and the COUP sample, before and after correction.}
  \label{fig:offsets}
 \end{center}
\end{figure*}

We wanted to check if a systematic offset in the positions of both catalogs was present. Thus, we calculated the median differences in the right ascension and declination of this first set. As can be seen in Figure \ref{fig:offsets}, we found  $\delta RA_{U-excess}$ = +0.274" and $\delta DEC_{U-excess}$ = +0.036", and $\delta RA_{H{\alpha}}$ = +0.228", and $\delta DEC_{H{\alpha}}$ = +0.026". We then corrected the COUP sources' right ascension and declination with these median offset values to refine the search of \textit{HST} counterparts, and did a new cross-match with these new coordinates. For the \textit{U-excess } sub-sample we found one less source, obtaining a sub-sample of 169 matches, whereas for the $H\alpha$ sub-sample we gained two more, obtaining 324 pairings.

As an extra control measure, we also performed the cross-match with the possibility of various pairings instead of just the nearest one. Not one of the \textit{HST} sources in our \textit{Manara2012} subsamples had more than one COUP counterpart nearby.
Table \ref{tab:offsets} shows the parameters from this procedure.

\section{Sample selection}
\label{sec:analysis}
\subsection{Spectral type and low mass}
To study the mass accretion rates and the X-ray luminosities as a function of $T_{eff}$, we only kept  sources with a known spectral type. This left us with 281 sources in the case of the $H\alpha$ subsample, and 164 in the \textit{U-excess} one. Spectral types were derived from \cite{Hillenbrand1997}, \cite{Luhman2000}, and \cite{Lucas2001}. For some sources, the spectral type was recalculated in \cite{Hillenbrand2013}. In these cases, we use the latest spectral type available. 

The X-ray emission of high-mass stars is non-coronal in origin, or comes from a different kind of corona, or originates in unresolved lower mass companions \citep{Flaccomio2012}. Because of this, we limited our study to stars with $M_*$ $<$ 2 $\rm M_{\odot}$. With this selection, the numbers decreased to 277 sources for the $H\alpha$ subsample and 163 for the \textit{U-excess} subsample.

Some sources did not have defined uncertainties for the mass accretion rates. We only used sources with valid mass accretion values and uncertainties, discarding the invalid ones. Thus, the \textit{U-excess} subsample was reduced to 154 sources. The $H\alpha$ remained the same.
This selection criteria is similar to that in \cite{Telleschi2007}. 

\subsection{Classical T Tauri stars and weak-line T Tauri stars}
The T Tauri family is divided into classical T Tauri stars (CTTS) and weak-line T Tauri stars (WTTS). This division takes into account the value of some accretion tracers to classify the objects into accretors or non-accretors. Several studies use this classification when studying the accretion properties of star-forming regions. For example,  \cite{Preibisch2005origin} and \cite{Flaccomio2003b}  use the $Ca_{II}$ infrared triplet lines ($\lambda$ = 8542\AA) equivalent width to classify their sample, as do \cite{Hillenbrand1998}; and in \cite{Telleschi2007} they use the $H\alpha$ line ($\lambda$ = 6563\AA), as in \cite{Gudel2007XESTa}. 

In our case, different classifications were possible. From the COUP data we accessed the values of the $Ca_{II}$ line for both of our subsamples. Furthermore, for the $H\alpha$ subsample, the $H\alpha$ line equivalent width was also available. Thus, a second classification analysis could be carried out for this subsample. 

Table \ref{tab:accretion_table} shows the mean accretion values for both subsamples. We note that, although both subsamples were classified using the $Ca_{II}$ method, not every object in them was included in the analysis. The classification method only selects objects with $EW(Ca_{II})$ $<$ -1\AA\ (CTTS) and $EW(Ca_{II})$ $>$ 1\AA\ (WTTS) \citep{Preibisch2005origin}. Thus, objects with -1\AA\ $<$ $EW(Ca_{II})$ $<$ 1\AA\ are not included in the classification. In the case of the $H\alpha$ subsample, we used 108 sources for these analyses, and for the \textit{U-excess} subsample we used 56 sources.

Several objects that we classified as WTTS have stronger accretion rate measurements than the mean value of the total CTTS, and several CTTS have lower accretion rate values than the mean WTTS value. As a result, the mean $\dot{M}_{acc}$ values of the WTTS and CTTS sources for both subsamples differ by less than 3-$\sigma$.

The excess emission in $H\alpha$ (and probably in \textit{U band}) present in WTTS may be due to cromospheric emission and not only to accretion. There have been studies that measure how strong this effect is and, in general, it is below the observed accretion rates \citep{Ingleby2011, Stelzer2013, Manara2013}. Accreting PMS with values of log$L_{acc}$ $\leq$ -3 log$L_{\odot}$ should be treated with caution because the line emission may be dominated by the contribution of chromospheric activity \citep{Manara2013a}. In our case, none of the WTTS with strong accretion measurements had log$L_{acc}$ values lower than -3 log$L_{\odot}$, so we assume no significant cromospheric dependence for their accretion-rate measurements.

Our data do not allow us to classify our subsamples between CTTS and WTTS clearly. Medium-high resolution spectra would be needed for that analysis. Thus, in this work, we do not use the classical and weak-lined T Tauri division.

\section{Relations between X-ray luminosity, stellar mass, and accretion rate}
\label{sec:correlations}
For the characterization of the X-ray properties, we used the following data from the COUP catalog: the total X-ray luminosity $L_X$ (corrected for interstellar absorption), which corresponds to the total 0.5 - 8.0 keV energy band, with units of \textit{erg $s^{-1}$}, and the equivalent width values of the $Ca_{II}$ line for the T Tauri classification, in units of \AA\ (see previous section). Importantly, \cite{Manara2012} formulate their analyses using a distance of 414 pc to the ONC, while the data in the COUP catalog assumes a distance of 450 pc. We corrected the latter with the distance of the former, using the equation used in the COUP catalog: L = 4$\pi$$D^2$F, with F being the X-ray flux and D being the distance in parsecs. We used a value of 0.39 dex for the uncertainty of $L_X$.

These X-ray parameters are selected as in \cite{Preibisch2005origin}. The main difference with that work is the spectral type classification \cite[we updated the spectral types of several objects with the information from][]{Hillenbrand2013} and the sample selection (we used only objects with M$_*$ $<$ 2 $\rm M_{\odot}$).

The rest of the stellar parameters were gathered from \cite{Manara2012}: the stellar masses $M_*$, in units of $\rm M_{\odot}$, were obtained from \cite{DAntona1994} models and the accretion rates, with units of $\rm M_{\odot}$$yr^{-1}$, were derived from that paper, using photometric data from the \textit{HST Treasury Program} (see Sect. ~\ref{sec:observations}).

In Table \ref{tab:coup_sample} we present the final catalog with all these parameters for both subsamples.

\subsection{Least-square fit and outlier rejection study}
In this work the regression analyses were carried out using two methods: a least-square approximation and an outlier rejection fit. 

For the first, we used a classical \textit{ordinary least square} (OLS) algorithm using the \textit{scipy's orthogonal distance regression} (ODR) package on \textit{Python} \citep{Jones2001, Boggs1990}. The ODR package allowed us to feed the uncertainties of our parameters into the analysis and take them into account to acquire the fit. Moreover, it also produced the errors of the fitted coefficients, which we used to compute the uncertainties in our results.

Because of the large spread of our data, we decided to include a Bayesian outlier rejection algorithm with \textit{Monte Carlo Markov Chain} (MCMC), using the \textit{astroML} package of \textit{Python} \citep{Vanderplas2012}. The points, which are identified by this method as not part of the fit with a probability greater than 99.7\% (3-$\sigma$ confidence), are selected as outliers. This method uses models that marginalizes over the probability that each point is an outlier. The MCMC is used for this marginalization process. Then, from the sets of maximum a posteriori likelihoods for the slope and intercept values, we  selected the median values for both parameters as the best-fit results.

This method relies heavily on the measurement uncertainty. A few good points might have been rejected if their uncertainties were underestimated. But since the sample and measurements come from the same homogeneous survey, these errors should not bias our result significantly.

The Spearman correlation analysis was used to test the relation between the parameters in our study, because it is less sensitive to outliers than other correlation analyses, such as Kendall $\tau$ or Pearson $\rho$.

\subsection{Stellar mass dependencies}
Many studies have shown that there is an inherent relation between the mass of the young star and its X-ray emission \citep{Flaccomio2003b, Preibisch2005origin, Telleschi2007, Ercolano2014} and between the stellar mass and the mass accretion rates \citep{Calvet2004, Muzerolle2003, Muzerolle2005,  Mohanty2005, Natta2006, Manara2012, Alcala2014}. We want to test these results with our data.

\begin{figure*}[ht]
 \begin{center}
 \subfigure[H$\alpha$ sample.]{%
    \includegraphics[width=0.5\textwidth]{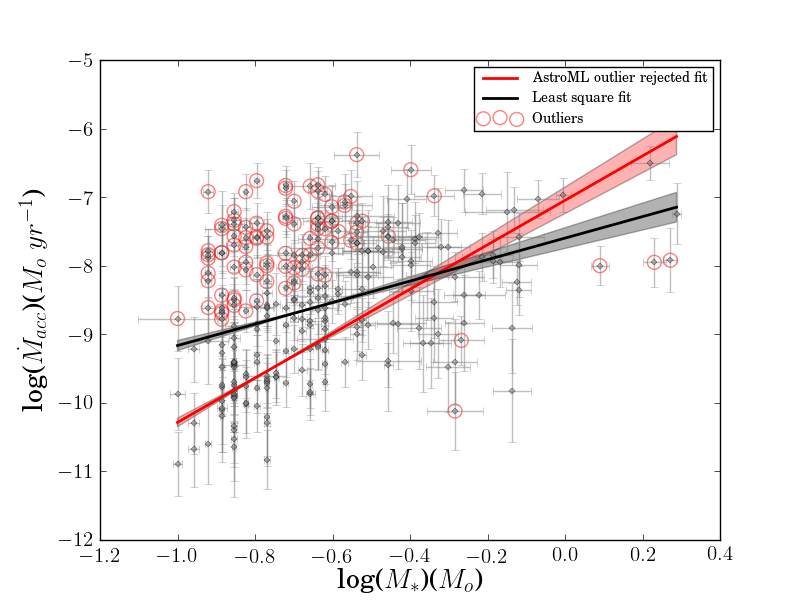}
  }%
  \subfigure[U-excess sample.]{%
    \includegraphics[width=0.5\textwidth]{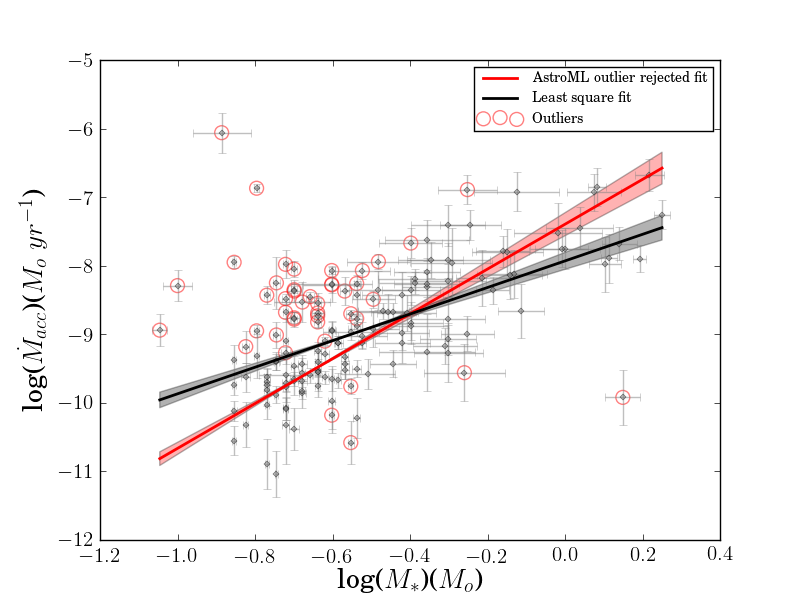}
      }%
 \caption{Stellar mass vs. mass accretion rates for both subsamples. The plot on the left corresponds to the \textit{H$\alpha$} subsample, and the one on the right to the \textit{U-excess} subsample. Regression lines (straight) are plotted with their respective errors in the slope (shaded areas) for the OLS regression analysis (black) and the outlier rejection regression method (red).} 
 \label{fig:mass_accretion}
 \end{center}
\end{figure*}

\textit{Stellar mass and accretion rates} - Figure \ref{fig:mass_accretion} and Table \ref{tab:regressions} show the correlation study between the mass of the star and its mass accretion rate. Important differences can be found for both subsamples. Strong correlation coefficients are found for the \textit{U-excess} subsample, with a relation of log$\dot{M}$ = (1.94$\pm$0.23)$\times$log$M_*$ - (7.95$\pm$0.12) using the minimum-square fit analysis. This result is consistent with  previous studies, log$\dot{M}$ = 2$\times$log$M_*$ - 7.5, which is used in the \cite{Telleschi2007} study of the TMC. But we can see how this is not the case for the $H\alpha$ subsample. The relation for this subsample is log$\dot{M}$ = (1.57$\pm$0.23)$\times$log$M_*$ - (7.60$\pm$0.15), but with weaker correlation coefficients. This discrepancy suggests that \textit{U-band excess} is a better accretion tracer than the $H\alpha$ line equivalent width \citep{Venuti2014}, and calls for a word of caution when using $H\alpha$ to derive accretion rates. In the case of the outlier rejection analysis, a steeper slope is found for the \textit{U-excess} subsample, log$\dot{M}$ = (3.29$\pm$0.25)$\times$log$M_*$ - (7.37$\pm$0.15), as well as for the $H\alpha$ subsample, (3.19$\pm$0.21)$\times$log$M_*$ - (7.09$\pm$0.16). In Sect. \ref{sec:discussion} we address this difference.

\begin{figure*}[ht]
 \begin{center}
 \subfigure[H$\alpha$ sample.]{%
    \includegraphics[width=0.5\textwidth]{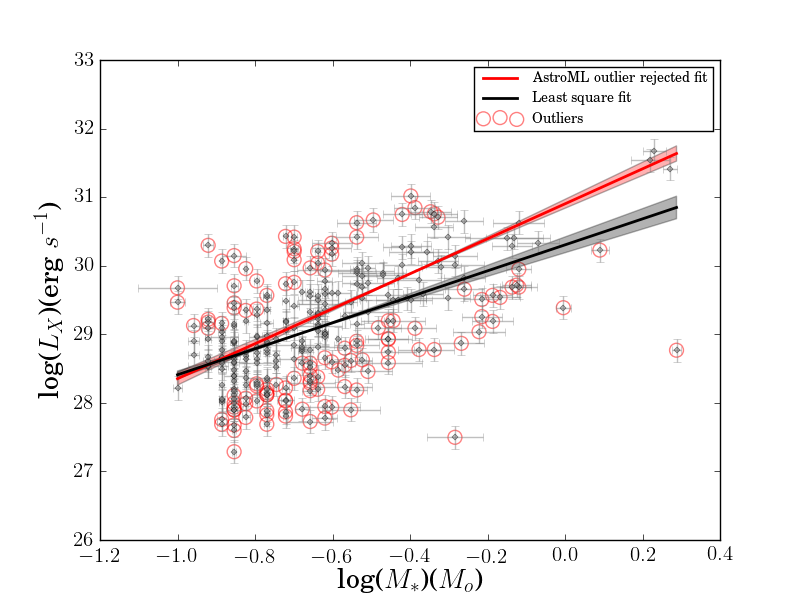}
  }%
  \subfigure[U-excess sample.]{%
    \includegraphics[width=0.5\textwidth]{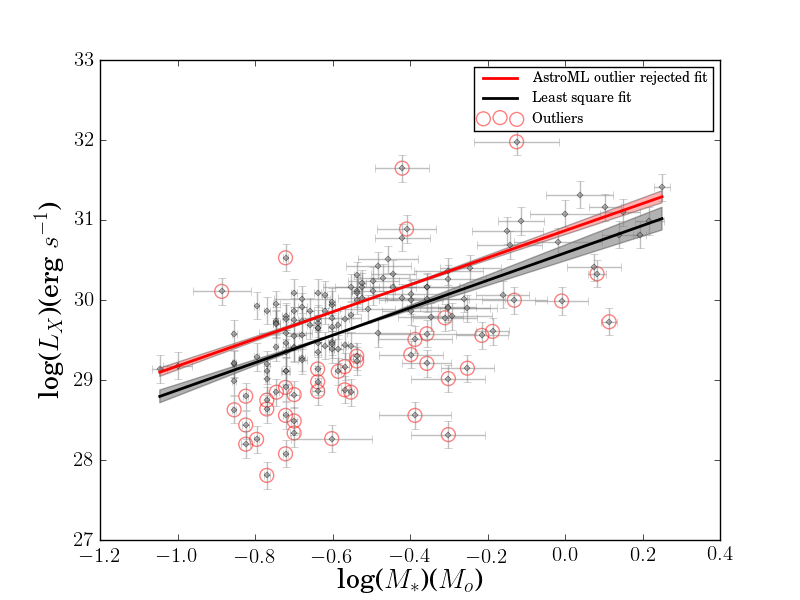}
      }%
\caption{X-ray luminosities vs. the mass of the sources. The plot on the left corresponds to the \textit{H$\alpha$} subsample, and the one on the right to the \textit{U-excess} subsample. Regression lines (straight) are plotted with their respective errors in the slope (shaded areas) for the OLS regression analysis (black) and the outlier rejection regression method (red). }
 \label{fig:lumX_mass}
\end{center}
\end{figure*}

\textit{Stellar mass and X-ray emission} - Figure \ref{fig:lumX_mass} shows the relations between the stellar mass and $L_X$, for both subsamples. The regression and correlation parameters of the different subsamples are summarized in Table \ref{tab:regressions}. For the \textit{U-excess} subsample, stronger correlations are present between these parameters than those found for the $H\alpha$ one, a similar result to that on the $\dot{M}_{acc}$ - $M_*$ study. In Sect. \ref{sec:discussion} we explain this critical difference.

\subsubsection{Selection bias}
\cite{Ercolano2014} study whether the $\dot{M}_{acc}$ - $M_*$ relation found for different samples of YSO can be a consequence of a selection bias or detection threshold. Apart for finding similar relations to our subsamples, they find that it is not affected by a selection bias.

The \textit{Manara2012} sample we used is almost complete down to the hydrogen-burning limit. Thus, we conclude that selection effects are not present in our $\dot{M}_{acc} - M$ relations. And given the X-ray detection limit of the \textit{Chandra} telescope, which is lower than our lowest X-ray luminosity value, we also deduce that our $L_X - M$ relations are also not affected by a selection bias owing to undetected sources.

\subsection{Accretion rate and X-ray luminosity}
We  now  focus on our primary objective: study the accretion processes of the PMS objects in the ONC and their relation with the X-ray emissions of the central stars. This interaction has also been studied in different works, such as \cite{Preibisch2005origin} and \cite{Drake2009} in Orion, or \cite{Telleschi2007} in Taurus. Here we want to reformulate the results obtained in these works, using different accretor tracers and measurements. In particular, we want to compare the analysis of \cite{Telleschi2007} in the TMC with our results in the ONC, and see if similar relations can be inferred.

In our work we are not interested in studying the differences between accretors and non-accretors. That classification is no longer appropriate for our study, as seen in Sect. \ref{sec:analysis}. This is why we  study the direct dependences and relations of the measured mass accretion rates and the X-ray luminosities. We take an approach similar to that of \cite{Telleschi2007}. Here they study the $L_X$ - $\dot{M}_{acc}$ relation, taking into account the inherent dependences of these properties with the mass of the star. Class II objects show a direct relation between $\dot{M}_{acc}$ and \textit{M}. Using different evolutionary tracks, \cite{Muzerolle2003, Muzerolle2005} found $\dot{M}$ $\propto$ $M^{2}$ and $\dot{M}$ $\propto$ $M^{2.1}$, respectively, so in \cite{Telleschi2007} they adopted the relation log$\dot{M}$ $\approx$ 2log\textit{M} - 7.5. They also compute a relation between log$L_X$ and log\textit{M} with their data. Finally, they relate the $L_X$ with the $\dot{M}_{acc}$ with the equation log$L_X$($\dot{M}$) = 0.85$\times$log($\dot{M})$ + 36.67.

Taking the same approach, we used the $\dot{M}_{acc}$ - $M_*$ and the $L_X$ - $M_*$ relations we found for both our subsamples and derived $L_X$ as a function of $\dot{M}_{acc}$. By using both the \textit{OLS} and the \textit{outlier detection and rejection} methods, different regression results and, consequently, $L_X(\dot{M}_{acc})$ equations, are obtained for each subsample. These results are shown in Table \ref{tab:lx_acc_equations}. With them, we computed the expected $L_X$ given the $\dot{M}_{acc}$ of the sources in our catalog. We called this our \textit{theoretical} $L_X(\dot{M}_{acc})$. This relation could define the $L_X$ - $\dot{M}_{acc}$ in Orion if both parameters were to only depend on the stellar mass. 

\begin{figure*}
\begin{tabular}{cc}
  \includegraphics[width=0.5\textwidth]{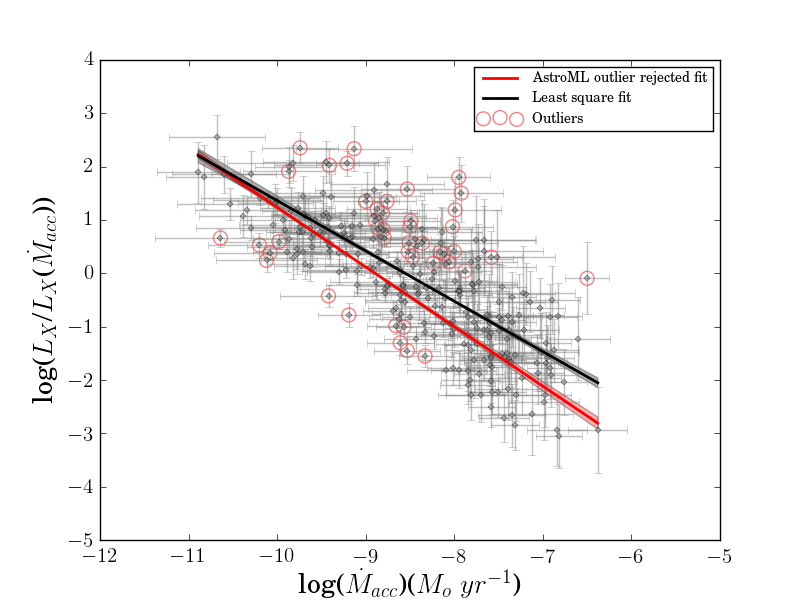}&\includegraphics[width=0.5\textwidth]{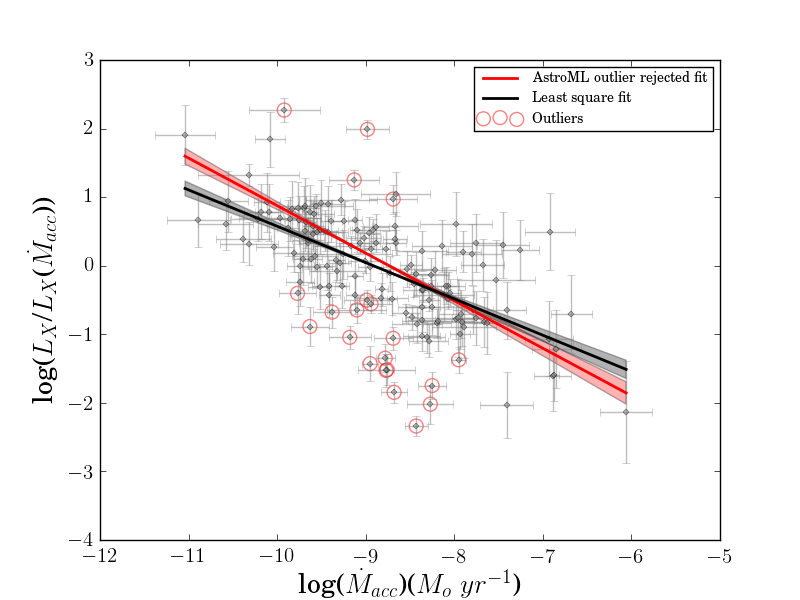}\\
  \includegraphics[width=0.5\textwidth]{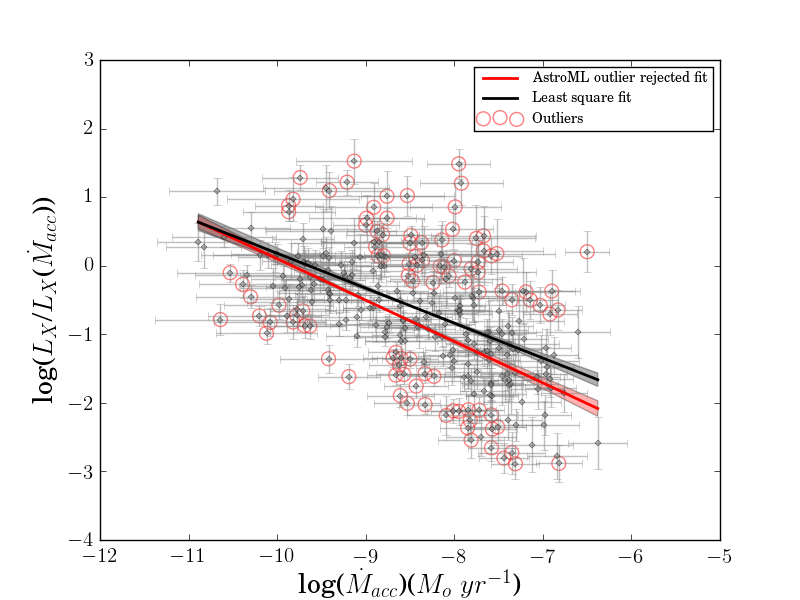}&\includegraphics[width=0.5\textwidth]{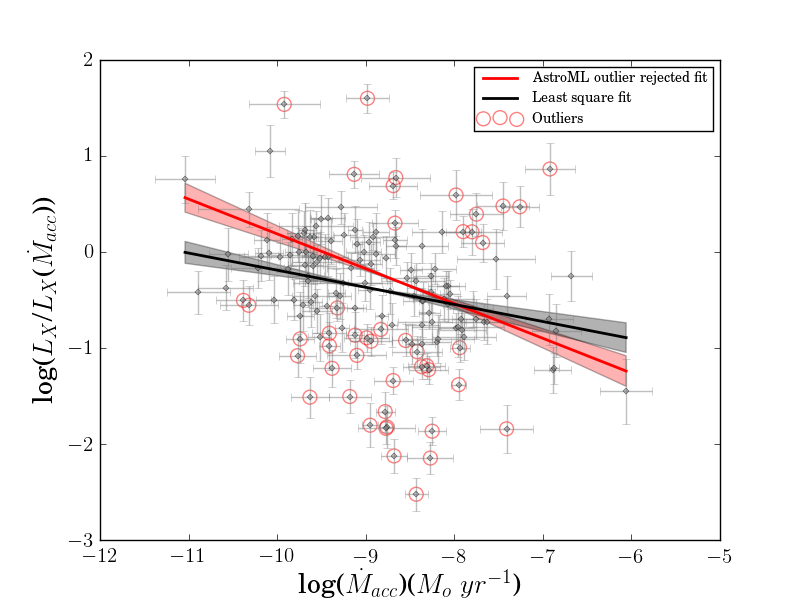}\\
\end{tabular}{}
\caption{Residual X-ray luminosity, after normalization with the \textit{M} - \textit{L$_{X}$} and \textit{M} - \textit{$\dot{M}$} relations, as a function of the mass accretion rate. Left figures correspond to the $H\alpha$ subsample. Right figures correspond to the \textit{U-excess} subsample. The top plots present the results using the OLS method for the initial regression analyses between $M_*$, $L_X$ and $\dot{M}_{acc}$. The bottom ones when using the outlier rejection method for those same analyses. Regression lines (straight), with their respective errors in the slope (shaded areas), are shown for the OLS (black) and outlier rejection (red) final regression results.}
 \label{fig:lumX_accretion}
\end{figure*}

Finally, we computed what we called the \textit{residual} $L_X$. This parameter was obtained by dividing the observed $L_X$ by the \textit{theorethical} $L_X$, given by our derived equations, for a given $\dot{M}_{acc}$. In Table \ref{tab:coup_sample} we include the different residual $L_X$ values of each source  that has been computed this way, for both subsamples. This residual X-ray luminosity compares the observed $L_X$ with the expected value if both the $L_X$ and the $\dot{M}_{acc}$ were to only depend on the stellar mass. It represents the excess (or deficit) of X-ray luminosities that are dependent on the $\dot{M}_{acc}$. In Figure \ref{fig:lumX_accretion} we plot this value against the $\dot{M}_{acc}$. If this ratio were  determined only by the $L_X$ - $M_*$ and $\dot{M}_{acc}$ - $M_*$ relations, then the values would scatter around a constant. As can be seen in the figure, they present a large scatter, pointing towards more dependencies, in addition to the stellar mass, for $L_X$ and $\dot{M}_{acc}$. 

Again, we used both the \textit{OLS} and the \textit{outlier detection and rejection} methods for these final plots. Table \ref{tab:regressions} summarizes all the regression results obtained for all the possible combinations of data and methods. The values under  \textit{"Least square algorithm"}  in Table \ref{tab:regressions} correspond to the regression results that were obtained using the OLS methodology, whilst the ones under  \textit{"Outlier rejection algorithm"}  correspond to the results obtained using the named method. The four top rows present the results of the regressions that were found directly from the data, whilst the bottom four show the regression results from the derived \textit{residual X-ray luminosity}. This last parameter is computed using what we defined here as the \textit{theoretical X-ray luminosity}, $L_X$($\dot{M}$), which in turn depends on the first relations. Thus, depending on the methodology used to derive the first relations on the top four rows (OLS or outlier rejection), different $L_X$($\dot{M}$) are obtained. This is represented in the four bottom rows. The top two present the results from the $L_X$($\dot{M}$) derived from the OLS methodology, while the bottom two represent the  $L_X$($\dot{M}$) derived from the outlier rejection analysis.

\section{Discussion}
\label{sec:discussion}
We will now discuss the correlations and trends found in the previous section. The main objective is to compare the residual X-ray luminosity - mass accretion rate relations from both Taurus, obtained in \cite{Telleschi2007}, and Orion, obtained in this work, which we do in Sect. \ref{sec:discuss_accretion}. Different methods yield different results, which we address in Sects. \ref{sec:ols_results} and \ref{sec:outlier_results}. In Sect. \ref{sec:interpretation}, we interpret these results.

\subsection{Comparing Taurus and Orion - OLS method regression results}
\label{sec:discuss_accretion}
\cite{Telleschi2007} find the following results when studying the X-ray emission differences between CTTS and WTTS: i) CTTS present smaller $L_X$ than WTTS, on average; ii) they also present smaller $L_X$/$L_*$; iii) and a correlation is found between the electron temperature and the total $L_X$ for WTTS, but not for CTTS (where the temperatures are higher than those in the WTTS). Their results point towards a magnetic coronal source for the X-ray emissions in TTS. Also, they suggest that the lower $L_X$ emission from CTTS is caused by coronal heating from an accretion disk and accretion processes. Denser accreting material from CTTS would cause a decreased heated plasma from the reconnection events, accounting for the lower measurements of $L_X$ in accreting systems than in WTTS. The cool plasma may also reorganize the magnetic loops after entering them, stretching and causing them not to reconnect (diminishing the X-ray emission therefore); or by just lowering their temperature because they are colder than them; or inducing radiative losses and more rapid cooling.

One of their main results regarding the accretion processes and the $L_X$ luminosities was the relation between the residual $L_X$
and the $\dot{M}_{acc}$ they they found for their CTTS sample.  The main difference to their work can be found in  the subsamples we  gathered for the Orion YSOs, which yield different results in each case.

\subsection{OLS method regression results}
\label{sec:ols_results}
As can be seen in Figures \ref{fig:mass_accretion} and \ref{fig:lumX_mass}, the spread in the data is considerable, over orders of magnitude. For the $H\alpha$ subsample, in fact, this spread is large enough s to produce ineffective correlation results as can be seen in the low Spearman $\rho$ coefficient for the $\dot{M}_{acc}$ - $M_*$ relation in Table \ref{tab:regressions}. This spread may be caused by  hidden variables that affect our data. 

The \textit{U-excess} subsample presents a stronger correlation between these parameters, pointing towards more coherent measurement of the mass accretion rates of the sources \citep{Venuti2014}. The process to derive the accretion rate, which usies the equivalent width of the $H\alpha$ line, is more complex than the one with the excess emission of the \textit{U band}. More factors enter into play in the former, leading to  more uncertainties in the final data, which maybe explains the larger spread in the plots for this subsample.

This is why we cannot assert the reliability of the rest of the results from the $H\alpha$ sample from this conclusion onward. The low $\rho$ coefficient in the $\dot{M}_{acc}$ - $M_*$ relation means the derived $L_X$ - $\dot{M}_{acc}$ one is not sufficiently reliable. Therefore, our analyses will focus on the \textit{U-excess} subsample from now on. Nevertheless, we report the results from the other subsample as a reference and when making a comparison with the \textit{U-excess} data.

The \textit{U-excess} subsample presents moderate/strong correlations between our four main parameters, $M_*$, $\dot{M}_{acc}$, $L_X$, and the residual $L_X$. In fact, this is very similar to those found in \cite{Telleschi2007}. The relation between $M_*$ and $\dot{M}_{acc}$, $\dot{M}_{acc}$ $\propto$ $M_*^{1.94 \pm 0.23}$ given by the \textit{least-square} method is totally compatible with that found in many previous studies. Also, the $L_X$ $\propto$ $M_*^{1.72 \pm 0.17}$ relation is compatible with that found in \cite{Preibisch2005origin} for the same region. We still find considerable spreads in the data but the results we find are consistent with those on Taurus in \cite{Telleschi2007}.

We find that the final relation between the residual X-ray luminosity and the mass accretion rate for the \textit{U-excess} subsample, $L_X$/ $L_X$($\dot{M}$) $\propto$ $\dot{M}^{-0.53 \pm 0.05}$, is very compatible with the relation found in Taurus in \cite{Telleschi2007},  $L_X$/ $L_X$($\dot{M}$) $\propto$ $\dot{M}^{-0.48 \pm 0.15}$.

\subsection{Outlier method regression results}
\label{sec:outlier_results}
To test the robustness of our fits, we used an \textit{outlier detection algorithm}. The problem with this method, as stated before, is that it is very sensitive to uncertainties in the dependent variable, in this case, $\dot{M}_{acc}$, $L_X$, and $L_X$/ $L_X$($\dot{M}$). Several sources present  sufficiently small errors as to be selected as outliers, although their position in the plot is consistent with the best-regression fit. Thus, the results obtained using this method should be taken cautiously.

The parameters we are working with (stellar mass, mass accretion rate, and X-ray luminosity) are not measured directly, but  obtained indirectly from other measurements (i.e., the equivalent width of the $H\alpha$ line or the excess in the \textit{U band} for the $\dot{M}_{acc}$, or the flux of the objects, which is obtained by measuring the photons that arrive to the X-ray detector, for the $L_X$). Consequently, the uncertainties should represent this process.

As  mentioned in the previous section, the relation obtained between the residual X-ray luminosity and the mass accretion rate using the \textit{OLS} method for the \textit{U-excess} subsample agrees considerably well with the results found by \cite{Telleschi2007} in Taurus. On the other hand, using the outliers detection method, the results on Orion differ considerably from those in Taurus. 

The relations obtained with the \textit{outliers detection} method between $M_*$ and $\dot{M}_{acc}$ in both subsamples have considerably steeper slopes  compared to those found in the literature: $\dot{M}_{acc}$ $\propto$ $M_*^{3.19 \pm 0.21}$ for the $H\alpha$ subsample and $\dot{M}_{acc}$ $\propto$ $M_*^{3.29 \pm 0.25}$ for the \textit{U-excess} one. This represents a difference of more than 3-$\sigma$ from the ones obtained using the \textit{OLS} method. In the case of the $L_X$ - $M_*$ relation, the results differ from one subsample to another. The \textit{U-excess} subsample presents compatible regression results between both methods, and between the relation found in \cite{Preibisch2005origin}. But the ones found for the $H\alpha$ subsample are considerably different, by a factor larger than 3-$\sigma$.

In Table \ref{tab:regressions} all the final $L_X$/$L_X(\dot{M}_{acc})$ -  $\dot{M}_{acc}$ relations are shown, and Figure \ref{fig:lumX_accretion} presents the final plots. Using both methods, we obtain eight different results: two plots for the $H\alpha$ subsample, each with two different regression results, depending on whether we used the \textit{OLS} or the \textit{outliers detection} method to derive the regressions between $L_X$ or $\dot{M}_{acc}$ and $M_*$ and between $L_X$/$L_X(\dot{M}_{acc})$ -  $\dot{M}_{acc}$; and the same in the case of the \textit{U-excess} subsample.

Out of these eight results, and taking into account the aforementioned issues with each subsample and methodology, we decided that only two of them were compatible with Taurus: the one from the \textit{U-excess} subsample that uses the \textit{OLS} method between the $M_*$ - $L_X$ and $M_*$ - $\dot{M}_{acc}$ relations and between the $L_X$/$L_X(\dot{M}_{acc})$ -  $\dot{M}_{acc}$ final relation, and the one from the $H\alpha$ subsample that uses the \textit{outliers detection} method for the initial relations and the \textit{OLS} analysis for the $L_X$/$L_X(\dot{M}_{acc})$ -  $\dot{M}_{acc}$ final relation.

The uncertainties shown in Table \ref{tab:regressions} for Orion are far narrower than those found in \cite{Telleschi2007} for Taurus. The 1-$\sigma$ range for the slope from \cite{Telleschi2007} is \textit{-0.33} to \textit{-0.63}, which is compatible with three more results from our analysis if we add the uncertainties from them (slopes \textit{-0.35 $\pm$ 0.07}, \textit{-0.67 $\pm$ 0.06,} and \textit{-0.68 $\pm$ 0.13}). With these, our final interpretation of the results, as stated in Section \ref{sec:interpretation}, becomes stronger. Nevertheless, we urge caution, given the issues we explained before.

One important result to point out is that the regression results obtained using the \textit{OLS} method are recovered using the \textit{outliers detection} method when the uncertainties on the dependent variables are increased sufficiently (1-$\sigma$, 2-$\sigma$, 3-$\sigma$...). 

The accretion processes present an important variability in time, which we  disregard. Applying an outliers detection and rejection method to these kind of data, without taking this characteristic into account may seem  risky, as some sources could be an order of magnitude off its quiescent value and be selected as outliers when, in fact, they  are not. This is another reason why the results of the outliers analysis must be taken with caution. Nevertheless, given the numbers in our data, this issue is not so critical.

\subsection{Results}
\label{sec:interpretation}
Our main result is the anti-correlation found between the residual $L_X$ and the $\dot{M}_{acc}$. We deduce two interpretations from these results.

The first  is that for higher accretion rates there is a deficit in the observed $L_X$, compared with the \textit{theoretical} one. This implies that, for stronger accretors, the observed $L_X$ is weaker. That is, the accretion processes seem to "cover" the emission of X-ray radiation from the YSOs. This result is also found in \citep{Telleschi2007}.

The second  is with regard to the scatter in the values on these plots. If the $L_X$ and $\dot{M}_{acc}$ are only dependent on the stellar mass of the YSOs, as we have assumed, these values would scatter around a constant value of one. But the higher scatter and the negative slope points towards other dependencies for these parameters (e.g., stellar rotation, Rossby number, magnetic field intensity). What is interesting is the similarity between the $L_X/L_X$($\dot{M}_{acc})$ - $\dot{M}_{acc}$ relation on Taurus and Orion. This suggests that the environment where the T Tauri stars are born is not one of the parameters that affects the $L_X/L_X$($\dot{M}_{acc})$ - $\dot{M}_{acc}$ relation. 

As a result, we can conclude that this anti-correlation between the residual $L_X$ and the mass accretion rate is independent of the region and environment; its arbitrary appearance both in the TMC and in the ONC, shows reasonably well how it is an inherent property of accreting systems. Similar analyses in other star-forming regions could reinforce this result.

\section {Conclusions}
\label{sec:conclussions}
We have gathered a sample of T Tauri stars in the Orion Nebula Cluster with known X-ray detections and mass accretion rates. Using the classification criteria in \cite{Telleschi2007}, we have divided our sample in CTTS and WTTS, finding that several WTTS have accretion rates comparable or even higher to those of  CTTS. Since our data do not allow us to confirm these classifications and the number of WTTS in this work is small compared to the whole catalog, we have used all the  subsamples for our analysis.

Two subsamples have been used in this work, depending on the method used to calculate the accretion rates of the sources. The $H\alpha$ sample used the equivalent width of this spectral line, and the \textit{U-excess} subsample, the color excess compared to the nominal values. 

Because of the large spread in our data, two regression analyses were carried out. First, we used a \textit{least-square} regression analysis, using the \textit{ODRPACK} of \textit{Scipy} on Python, obtaining results compatible with those in the literature and with \cite{Telleschi2007} for one of our subsamples. Then, we detected and rejected outliers from our samples using the \textit{astroML MCMC} package of Python. This method is very sensitive to the uncertainties from the dependent variables and some sources were rejected as outliers as a result of too small error values, which thus affect the final regression-fit results. Using this method, only one more result was compatible with that from Taurus in \cite{Telleschi2007}. By enlarging the uncertainties in the dependent variables, the \textit{OLS} regression results were recovered.

We find that a clear anti-correlation between the residual X-ray luminosity and the accretion rates is present in both subsamples, although we cannot assert the reliability of the $H\alpha$ results because of the weak correlation found between its previous parameters. In the case of the \textit{U-excess} subsample, our results are consistent with those found in the Taurus Molecular Cloud. This result, independent of the environment studied, points towards an inherent property in these type of YSOs; the accretion processes causes a decrease in the emission of X-rays. Different theories have been postulated for this situation, such as thermalization by the X-rays of the stream being accreted, which causes a reduction in the X-ray measurements.

\bibliographystyle{aa}
\bibliography{biblio}

\begin{acknowledgements}
{This work has been made possible thanks to the ESAC Science Operations Division research funds,  code EXPRO IPL-PSS/GP/gp/44.2014, plus support from the ESAC Space Science Faculty. Pablo Riviere-Marichalar and Carlo Manara acknowledge funding from the ESA Research Fellowship program. H. Bouy is funded by the Spanish Ramón y Cajal fellowship program number RYC-2009- 04497. This research made use of the SIMBAD database, operated at the CDS, Strasbourg, France. This work made  extensive use of Topcat (TOPCAT http:// www.star.bristol.ac.uk/~mbt/topcat/ and STILTS, Taylor 2005, 2006). We thank the referee, Manuel G\"udel, for his comments and corrections, which help to make this work complete. We would also like to thank Benjamin Montesinos and Nuria Huelamo for their comments and help.}
\end{acknowledgements}

\newpage

\begin{table*}[htbp]
\centering
\caption{Key parameter comparison between the TMC and the ONC.}
\begin{tabular}[width=0.5\textwidth]{lc c c c c c c c cl}
\hline\hline\\
Region&X-ray Program\tablefootmark{1} & Distance (pc)\tablefootmark{2} & Age (Myr)\tablefootmark{3}  & $N_{total}$\tablefootmark{4} \\
\hline
Orion Nebula Cluster & COUP & 414 & $\sim$ 2 & 277/163\\ 
Taurus Molecular Cloud & XEST & 150 & $\sim$ 1-2 & 105\\ 
\hline
\label{tab:tmc_onc_comparison}
\end{tabular}
\tablefoot{
\tablefootmark{1}{Name of the program from where the data were extracted (ONC Ref. \cite{Preibisch2005origin};TMC Ref. \cite{Telleschi2007, Gudel2007XESTa}).}\\
\tablefootmark{2}{Distance to the star-forming region (ONC Ref. \cite{Menten2007}; TMC Ref. \cite{Telleschi2007}).}\\
%\tablefootmark{3}{Total stellar mass (ONC Ref. data from \cite{DaRio2012}; TMC Ref. \cite{}).}\\
\tablefootmark{3}{Mean log stellar age (ONC Ref. data from COUP and \cite{Reggiani2011}; TMC Ref. \cite{Telleschi2007}).}\\
\tablefootmark{4}{Number of sources used in the TMC analyses (Ref. \cite{Telleschi2007}) and in the analyses of the ONC. (Ref. \cite{Preibisch2005origin}, this wok). In the case of the ONC, the different numbers correspond to the sizes of each subsample studied.}\\
}
\end{table*}

\begin{table*}
\centering
\caption{Cross-matching coordinates and parameters before and after the offset corrections.}
\begin{tabular}[width=0.5\textwidth]{@{\extracolsep{4pt}}c c c c c c c@{}}
\hline\hline\\
&Mean $\Delta$(")\tablefootmark{1} & Mean$\Delta$$\alpha$(")\tablefootmark{2} & Mean$\Delta$$\delta$(")\tablefootmark{2}&Mean $\Delta$(")\tablefootmark{1} & Mean$\Delta$$\alpha$(")\tablefootmark{2} & Mean$\Delta$$\delta$(")\tablefootmark{2} \\
\cline{2-4} \cline{5-7}
(Subsample) & \multicolumn{3}{c}{(\textit{Halpha})} & \multicolumn{3}{c}{(\textit{U-excess})} \\
\hline
Before correction&0.320&+0.248&+0.026&0.337&+0.274&+0.036\\ 
After correction&0.208&+0.130e-3&+0.011&0.199&+0.007&+0.003\\ 
\hline
\label{tab:offsets}
\end{tabular}
\tablefoot{
\tablefootmark{1}{Mean position offset between X-ray COUP sources and both \textit{HST} \cite{Manara2012} subsamples.}\\
\tablefootmark{2}{Mean right ascension and declination offsets for both sub-samples.}\\
}
\end{table*}

\begin{comment}
\begin{table*}[htbp]
\centering
\caption{CTTS and WTTS classification results of both sub-samples. The $Ca_{II}$ tracer was available for both sub-samples, but the $H{\alpha}$ one only for the $H{\alpha}$ sub-sample.}
\begin{tabular}[width=0.5\textwidth]{@{\extracolsep{4pt}}c c c c c@{}}
\hline\hline\\
Classification Method & $H\alpha$ sub-sample CTTS  & $H\alpha$sub-sample WTTS & \textit{U-excess} sub-sample CTTS & \textit{U-excess} sub-sample WTTS\\
\hline\\
$Ca_{II}$ method & 69 ($\sim$25\%)& 39 ($\sim$14\%)&14 ($\sim$9\%) & 42 ($\sim$27\%)\\
H$\alpha$ method & 212 ($\sim$77\%)& 56 ($\sim$20\%)& - & - \\
\textit{Barrado} method & 220 ($\sim$79\%)& 51 ($\sim$18\%)& - & - \\
\hline
\label{tab:ttauri_family}
\end{tabular}
\end{table*}
\end{comment}

\begin{table*}[htbp]
\centering
\caption{Mean values of log($\dot{M}_{acc}$) for both CTTS and WTTS classified with the different accretion tracers. The standard deviation is used as the uncertainty.}
\begin{tabular}{@{\extracolsep{4pt}}c c c c c@{}}
\hline\hline\\
& $Mean_{\dot{M}_{acc}}$ CTTS $\pm$ $\sigma$  & $Mean_{\dot{M}_{acc}}$ WTTS $\pm$ $\sigma$ & $Mean_{\dot{M}_{acc}}$ CTTS $\pm$ $\sigma$ & $Mean_{\dot{M}_{acc}}$ WTTS $\pm$ $\sigma$\\
\cline{2-3} \cline{4-5}
(Subsample)& \multicolumn{2}{c}{(\textit{Halpha})}&\multicolumn{2}{c}{(\textit{U-excess})} \\
\hline
$Ca_{II}$ method & - 7.99 $\pm$ 0.94 & - 8.84 $\pm$ 0.98 &- 7.78 $\pm$ 0.80 & - 9.00 $\pm$ 0.80 \\
H$\alpha$ method& - 8.11 $\pm$ 0.79 & - 9.56 $\pm$ 0.68  & - & - \\
\textit{Barrado} method& - 8.30 $\pm$ 0.91 & - 9.00 $\pm$ 1.01 & - & - \\
\hline
\label{tab:accretion_table}
\end{tabular}
\end{table*}

\begin{table*}[htbp] 
\caption{X-ray and stellar parameter information for both subsamples.} 
\scriptsize\centering 
\begin{tabular}[width=1\textwidth]{lc c c c c c c c c c c cl} 
\hline \hline 
Coup id\tablefootmark{1} & $RA_{COUP}$\tablefootmark{2} & $DEC_{COUP}$\tablefootmark{2} & log($L_X$)\tablefootmark{3} & OM id\tablefootmark{1} & $RA_{Manara}$\tablefootmark{2} & $DEC_{Manara}$\tablefootmark{2} & log($M_*$) & log($\dot{M}_{acc})$  & SpT & log($L_X$/$L_X$($\dot{M}_{acc}$))\tablefootmark{4} & log($L_X$/$L_X$($\dot{M}_{acc}$))\tablefootmark{4}\\ 
& (deg) & (deg) & (erg $s^{-1}$) & & (deg) & (deg) & ($M_O$) & ($M_O$ $yr^{-1}$) & & \textit{OLS} method & \textit{Outlier} method\\ 
\hline 
\multicolumn{11}{c}{{$H{\alpha}$}}\\ 
\hline 
11&83.670400&-5.378410&30.29$\pm$0.17&101&83.670246&-5.378431&-0.40$\pm$0.05&-6.60$\pm$0.36&M1&-1.22$\pm$0.78 & -1.05$\pm$0.42\\ 
14&83.673400&-5.399380&28.69$\pm$0.17&108&83.673392&-5.399303&-0.82$\pm$0.01&-10.02$\pm$0.43&M5&1.31$\pm$0.37 & 0.21$\pm$0.15\\ 
16&83.674400&-5.363780&28.53$\pm$0.17&117&83.674233&-5.363750&-0.92$\pm$0.01&-10.60$\pm$0.59&M7&1.85$\pm$0.34 & 0.54$\pm$0.18\\ 
29&83.694200&-5.390450&29.75$\pm$0.17&144&83.693996&-5.390442&-0.51$\pm$0.07&-7.78$\pm$0.36&M2&-0.34$\pm$0.45 & -0.60$\pm$0.30\\ 
37&83.699800&-5.394550&28.03$\pm$0.17&153&83.699687&-5.394536&-0.80$\pm$0.01&-9.78$\pm$0.36&M5&0.36$\pm$0.38 & -0.65$\pm$0.18\\ 
28&83.693400&-5.408840&30.79$\pm$0.17&157&83.693308&-5.408869&-0.35$\pm$0.05&-9.13$\pm$0.66&M0&2.34$\pm$0.42 & 1.57$\pm$0.40\\ 
40&83.700700&-5.377370&28.90$\pm$0.17&158&83.700454&-5.377372&-0.89$\pm$0.01&-9.77$\pm$0.42&M3&1.22$\pm$0.31 & 0.21$\pm$0.13\\ 
117&83.737300&-5.368430&29.53$\pm$0.17&185&83.737183&-5.368425&-0.52$\pm$0.01&-8.79$\pm$0.32&M3&0.67$\pm$0.15 & 0.02$\pm$0.15\\ 
133&83.746700&-5.385510&28.78$\pm$0.17&194&83.746587&-5.385497&-0.89$\pm$0.01&-9.48$\pm$0.47&M4&0.75$\pm$0.16 & -0.15$\pm$0.21\\ 
137&83.748200&-5.400080&29.15$\pm$0.17&199&83.748138&-5.400053&-0.64$\pm$0.01&-8.32$\pm$0.30&M4&-0.28$\pm$0.22 & -0.75$\pm$0.18\\ 
...\\ 
\hline 
\multicolumn{11}{c}{\textit{U-excess}}\\ 
\hline 
1&83.622700&-5.393730&29.74$\pm$0.17&70&83.622692&-5.393733&-0.74$\pm$0.00&-11.04$\pm$0.34&M5&1.90$\pm$0.44 & 0.77$\pm$0.27\\ 
43&83.703500&-5.388330&30.24$\pm$0.17&146&83.703458&-5.388329&-0.49$\pm$0.07&-8.92$\pm$0.26&M1&0.53$\pm$0.16 & 0.16$\pm$0.15\\ 
55&83.710600&-5.393160&29.17$\pm$0.17&167&83.710400&-5.393148&-0.57$\pm$0.01&-8.37$\pm$0.20&M1&-1.03$\pm$0.22 & -1.20$\pm$0.14\\ 
89&83.726200&-5.359870&29.59$\pm$0.17&174&83.726037&-5.359843&-0.48$\pm$0.08&-7.94$\pm$0.11&M0&-0.99$\pm$0.22 & -1.00$\pm$0.16\\ 
65&83.716800&-5.405220&29.46$\pm$0.17&175&83.716704&-5.405229&-0.64$\pm$0.01&-8.82$\pm$0.08&M3&-0.34$\pm$0.24 & -0.67$\pm$0.24\\ 
71&83.719200&-5.401070&30.17$\pm$0.17&177&83.719104&-5.401034&-0.44$\pm$0.06&-9.43$\pm$0.20&M1&0.91$\pm$0.25 & 0.36$\pm$0.22\\ 
107&83.733300&-5.386970&31.32$\pm$0.17&178&83.733208&-5.386923&0.04$\pm$0.09&-7.45$\pm$0.30&K2&0.31$\pm$0.48 & 0.47$\pm$0.24\\ 
118&83.737600&-5.383360&28.82$\pm$0.17&189&83.737512&-5.383354&-0.70$\pm$0.01&-10.38$\pm$0.31&M4&0.40$\pm$0.34 & -0.49$\pm$0.24\\ 
122&83.740800&-5.380900&30.09$\pm$0.17&191&83.740746&-5.380872&-0.70$\pm$0.01&-8.05$\pm$0.11&M3&-0.39$\pm$0.20 & -0.44$\pm$0.16\\ 
228&83.770100&-5.377450&28.34$\pm$0.17&224&83.770043&-5.377404&-0.70$\pm$0.01&-8.76$\pm$0.08&M4&-1.52$\pm$0.23 & -1.82$\pm$0.24\\ 
...\\ 
\hline 
...\\ 
\hline 
\hline 
\label{tab:coup_sample}  
\end{tabular}  
\tablefoot{ 
\tablefootmark{1}{Source ids as in the COUP catalog and in the \cite{Manara2012} samples.}\  
\tablefootmark{2}{Source coordinates as in the COUP catalog and in the \cite{Manara2012} samples.}\ 
\tablefootmark{3}{Logarithmic value of the X-ray luminosity, computed at a distance of 414pc.}\ 
\tablefootmark{4}{Residual X-ray luminosities, computed using the regression results from the        extit{OLS} method in the first case and from the \textit{outlier detection and rejection} algorithm in the second.}\ 
} 
\end{table*}

\begin{table*}[htbp] 
\caption{Summary of results of the different correlations analyses. The relations correspond to the logarithmic equation of the parameters.}
\centering 
\begin{tabular}{@{\extracolsep{4pt}}c c c c c c c c c@{}}
\hline\hline 
Correlation\tablefootmark{1} & Subsample\tablefootmark{2} & N\tablefootmark{3} & \multicolumn{2}{c}{Least square algorithm}\tablefootmark{4} & \multicolumn{2}{c}{Outlier rejection algorithm}\tablefootmark{4} & \multicolumn{2}{c}{Spearman Parameters} \\  
\cline{4-5} \cline{6-7}\cline{8-9}
& & &a & b & a & b & $\rho$\tablefootmark{5} & g\tablefootmark{6}\\ 
\hline 
$\dot{M}$ vs. \textit{M}&H$\alpha$&277&1.57 $\pm$ 0.23&-7.60 $\pm$ 0.15&3.19 $\pm$ 0.21&-7.09 $\pm$ 0.16&0.38&Weak\\ 
$\dot{M}$ vs. \textit{M}&\textit{U-excess}&154&1.94 $\pm$ 0.23&-7.93 $\pm$ 0.12&3.29 $\pm$ 0.25&-7.37 $\pm$ 0.15&0.55&Moderate\\ 
\hline 
$L_X$ vs. \textit{M} &H$\alpha$&277&1.90 $\pm$ 0.17&30.33 $\pm$ 0.12&2.72 $\pm$ 0.17&30.89 $\pm$ 0.09&0.52&Moderate\\ 
$L_X$ vs. \textit{M} &\textit{U-excess}&154&1.72 $\pm$ 0.17&30.59 $\pm$ 0.10&1.75 $\pm$ 0.12&30.89 $\pm$ 0.07&0.61&Strong\\ 
\hline
 \multicolumn{9}{c}{Residual $L_X$ determined using minimum-square fit analysis}\tablefootmark{7}\\ 
\hline
$L_X$/ $L_X$($\dot{M}$) vs. \textit{$\dot{M}$}&H$\alpha$&277&-0.94 $\pm$ 0.05&-8.07 $\pm$ 0.40&-1.09 $\pm$ 0.04&-9.65 $\pm$ 0.39&-0.78&Strong\\ 
$L_X$/ $L_X$($\dot{M}$) vs. \textit{$\dot{M}$}&\textit{U-excess}&154&-0.53 $\pm$ 0.05&-4.72 $\pm$ 0.43&-0.67 $\pm$ 0.06&-5.85 $\pm$ 0.53&-0.63&Strong\\ 
\hline 
 \multicolumn{9}{c}{Residual $L_X$ determined using MCMC outlier rejection fit analysis}\tablefootmark{7}\\ 
\hline
$L_X$/ $L_X$($\dot{M}$) vs. \textit{$\dot{M}$}&H$\alpha$&277&-0.59 $\pm$ 0.05&-5.50 $\pm$ 0.43&-0.68 $\pm$ 0.13&-6.51 $\pm$ 1.22&-0.55&Moderate\\ 
$L_X$/ $L_X$($\dot{M}$) vs. \textit{$\dot{M}$}&\textit{U-excess}&154&-0.19 $\pm$ 0.05&-2.04 $\pm$ 0.47&-0.35 $\pm$ 0.07&-3.28 $\pm$ 0.58&-0.29&Weak\\ 
\hline 
\label{tab:regressions} 
\end{tabular} 
\tablefoot{
\tablefootmark{1}{Parameters to analyze.}\\
\tablefootmark{2}{Subsample from which the data are referred.}\\
\tablefootmark{3}{Number of objects in the subsample.}\\
\tablefootmark{4}{\textit{a} stands for the slope of the linear regression and \textit{b} for the intercept. Errors for the least square analyses are the standard deviation as given in the ODR algorithm. Errors for the outlier analyses are the standard deviation of all the values computed in the MCMC algorithm for each parameter.}\\
\tablefootmark{5}{Spearman's $\rho$ rank correlation coefficient.}\\
\tablefootmark{6}{Goodness-of-fit as given by $\rho$.}\\
\tablefootmark{7}{Different regression methods between $M_*$, $L_X$ and $\dot{M}_{acc}$ yield different $L_X$($\dot{M}$) equations. Rows five and six show the regression results between $L_X$/$L_X$($\dot{M}$) and $\dot{M}_{acc}$ when the OLS method was used for the first regressions, while rows seven and eight, when the outlier rejection methodology was applied.}\\
}
\end{table*} 

\begin{table*}[htbp] 
\centering 
\caption{$L_X$ vs. $\dot{M}_{acc}$ equations for each subsample. Each one is computed using the regression results of $L_X$, $M_*$ and $\dot{M}_{acc}$. The ones using the results from the minimum-square analysis are shown on top, and the ones using the outlier detection and rejection analysis are shown below.} 
\begin{tabular}[width=0.5       extwidth]{lc c c c c cl} 
\hline\hline 
Sample & $L_X$ - $\dot{M}_{acc}$\\ 
\hline
\multicolumn{2}{c}{Using minimum-square fit analysis}\\ 
\hline
$H\alpha$ & log$L_X$ = (1.21$\pm$0.28)log$\dot{M}$ + (39.49$\pm$2.09)\\ 
\textit{U-excess} & log$L_X$ = (0.89$\pm$0.19)log$\dot{M}$ + (37.60$\pm$1.51)\\ 
\hline 
\multicolumn{2}{c}{Using MCMC outlier rejection fit analysis}\\ 
\hline
$H\alpha$ & log$L_X$ = (0.85$\pm$0.11)log$\dot{M}$ + (36.93$\pm$0.74)\\ 
\textit{U-excess} & log$L_X$ = (0.53$\pm$0.08)log$\dot{M}$ + (34.81$\pm$0.55)\\ 
\hline 
\label{tab:lx_acc_equations} 
\end{tabular} 
\end{table*} 

\end{document}